\documentclass[fleqn,twoside]{article}
\usepackage{espcrc2}

\usepackage{graphicx}
\usepackage[figuresright]{rotating}

\newcommand{\AmS}{{\protect\the\textfont2
  A\kern-.1667em\lower.5ex\hbox{M}\kern-.125emS}}

\title{Outflows from Quasars and Ultraluminous X--ray Sources}

\author{A.R. King\address{Theoretical Astrophysics Group, University
        of Leicester, Leicester LE1 7RH U.K.}}

\def\msun{{\rm M_{\odot}}}

\def\be{\begin{equation}}
\def\ee{\end{equation}}

\def\le{{L_{\rm Edd}}}
\def\msun{{\rm M_{\odot}}}

\def\rp{{R_{\rm ph}}}
\def\rs{{R_{\rm s}}}
\def\mo{{\dot M_{\rm out}}}
\def\me{{\dot M_{\rm Edd}}}
       
\begin{document}

\begin{abstract}
Mass outflows from Eddington--limited accreting compact objects appear
to be a very widespread phenomenon. They may provide the soft excess
observed in quasars and ULXs, and imply that such objects have a major
effect on their surroundings. In particular they allow a simple
parameter--free argument for the $M_{\rm BH} - \sigma$ relation for
galaxies, and offer a straightforward interpretation of the emission
nebulae seen around ULXs.

\vspace{1pc}
\end{abstract}

\maketitle

\section{INTRODUCTION}

Recent {\it XMM--Newton} observations of bright quasars (Pounds et
al., 2003a, b; Reeves et al., 2003) give strong evidence for intense
outflows from the nucleus, with mass rates $\mo \sim 1\msun~{\rm
yr}^{-1}$ and velocity $v \sim 0.1c$, in the form of blueshifted
X--ray absorption lines. 
These outflows closely resemble those
recently inferred in a set of ultraluminous X--ray sources (ULXs) with
extremely soft spectral components (Mukai et al, 2003; Fabbiano et
al., 2003).
Simple theory shows that the outflows are
probably optically thick to electron scattering, with a photosphere of
$\sim 100$ Schwarzschild radii, and driven by
continuum radiation pressure. In all cases the outflow velocity is
close to the escape velocity from the scattering photosphere. As a
result the outflow momentum flux is comparable
to that in the Eddington--limited radiation field, i.e.
\begin{equation}
\mo v \simeq {\le\over c},
\label{mom}
\end{equation}
where $\mo$ is the mass outflow rate and $\le$ the Eddington
luminosity, while the mechanical energy flux is
\begin{equation}
{1\over 2}\mo v^2 \simeq {\le^2\over 2\mo c^2}. 
\label{en}
\end{equation}

These outflows appear to be a widespread phenomenon, not only in
currently observed systems such as quasars and ULXs, but also in the
growth of supermassive black holes in the centres of galaxies in the
past.

\section{OUTFLOWS FROM EDDINGTON--LIMITED ACCRETORS}

I outline here a simple theory of outflows from black holes accreting
at rates comparable with the Eddington value
\begin{equation}
\dot M_{\rm Edd} = {4\pi GM\over \eta \kappa c}.
\label{medd}
\end{equation}
Here $\eta c^2$ is the accretion yield from unit mass, and
$\kappa$ is the electron scattering opacity. 
We assume that the outflow is radial, in a double cone 
occupying solid angle $4\pi b$,
and has constant speed $v$ for sufficiently large radial distance
$r$. I will justify the second assumption later in this Section.
Mass conservation implies an outflow density
\begin{equation}
\rho = {\mo\over 4\pi vbr^2}.
\label{cons}
\end{equation}
%The nature of the outflow depends on $b$. If $b \sim 1$ we can neglect
%scattering of photons from the sides of the outflow, while for $b <<
%1$ this process is dominant. For completeness we first briefly revisit
%the case $b \sim 1$ (cf Pounds et al. 2003) 

The electron scattering
optical depth through the outflow, viewed from infinity down to radius
$R$, is
\begin{equation}
\tau = \int_R^\infty\kappa\rho{\rm d}r = {\kappa\mo\over 4\pi vbR}.
\label{tau}
\end{equation}
From (\ref{medd}, \ref{tau}) we get
\begin{equation}
\tau = {1\over 2\eta b}{R_{\rm s}\over R}{c\over v}{\mo\over
\dot M_{\rm Edd}}.
\end{equation}
Defining the photospheric radius $R_{\rm ph}$ as the point $\tau = 1$ gives
\begin{equation}
{R_{\rm ph}\over R_{\rm s}} = {1\over 2\eta b}{c\over v}{\mo\over \dot
M_{\rm Edd}}\simeq {5\over b}{c\over v}{\mo\over \dot M_{\rm Edd}}
\label{phot}
\end{equation}
where we have taken $\eta \simeq 0.1$ at the last step. If $b \leq
1, v/c < 1$ we see that $R_{\rm ph} > R_{\rm s}$ for any outflow rate
$\mo$ of order $\dot M_{\rm Edd}$, that is, such outflows are
Compton thick. If instead $b << 1$, photons typically escape from the side of
the outflow rather than making their way radially outwards through all
of it. Almost all of the photons escape in this way within 
radial distance $r = R_{\perp}$ 
where the optical depth across the flow
\begin{equation}
\tau_{\perp} \simeq \kappa\rho(r) b^{1/2}r
\label{tauperp}
\end{equation}
is of order unity. Thus
\begin{equation}
{R_{\perp}\over R_{\rm s}} = {1\over 2\eta b^{1/2}}{c\over v}{\mo\over \dot
M_{\rm Edd}}\simeq {5\over b^{1/2}}{c\over v}{\mo\over \dot M_{\rm Edd}},
\label{rperp}
\end{equation}
and we again conclude that the outflow is Compton thick for $\mo
\sim \me$.

This evidently implies that much of the emission from such
objects will be thermalized and observed as a softer spectral
component (see eq \ref{teff} below). The observed harder X--rays must
presumably either be produced near the skin of the outflow (i.e. at
moderate $\tau$), or result from shocks within the outflow. In both
cases their total luminosity must be lower than that of the
thermalized soft component.

We now investigate how the outflow is driven.
Since the wind is Compton thick most of the photons have scattered
and thus on average given up their original momentum to the outflow.
Outside the radius $R_{\rm ph}$ or $R_{\perp}$ the photons decouple
from the matter and there is no more acceleration. This justifies our
assumption that $v$ is constant for large $r$, and it is
self--consistent to use the assumption to integrate inwards to 
$R_{\rm ph}$ or $R_{\perp}$. 

To ensure that the matter
reaches the escape speed we assume that the radii $R_{\rm
ph}, R_{\perp}$ to lie close to the escape radius $R_{\rm esc}$, i.e.
\begin{equation}
R_{\rm ph}, R_{\rm \perp} \simeq R_{\rm esc} = {c^2\over v^2}R_{\rm s}.
\label{esc}
\end{equation}
If instead the photosphere is outside the escape radius, i.e. there is
substantial photon trapping, one would insert an optical depth factor
$\tau$ on the rhs. The quasar observations referred to above, and the
normalization of the $M_{\rm BH} - \sigma$ relation derived below both
suggest that $\tau \simeq 1$, and I adopt this value in the rest of
this article.

From (\ref{esc}) and (\ref{phot}, \ref{rperp}) we find
\begin{equation}
{v\over c} \simeq {2\eta f\me\over \mo},\ \  R_{\rm ph,\ \perp}
\simeq \biggr({\mo\over 2\eta f\me}\biggr)^2
\label{scaling}
\end{equation}
where $f = b, b^{1/2}$ in the two cases $b < 1, b << 1$. 
We can write these formulae more compactly as
\begin{equation}
{v\over c} = {2f\le\over \mo c^2}
\label{v}
\end{equation}
\begin{equation}
{R_{\rm ph,\ \perp}\over \rs} = \biggl[{\mo c^2\over 2f\le}\biggr]^2,
\label{r}
\end{equation}
We note
that $f \sim 1$ except for very narrowly collimated outflows ($b
< 10^{-2}$).

An immediate consequence of (\ref{v}) is 
\begin{equation}
\mo v = 2f{\le\over c},
\label{mom}
\end{equation}
i.e. the momentum flux in the wind is always of the same order as that
in the Eddington--limited radiation field, as expected for an
Compton thick wind driven by radiation pressure. The energy flux
(mechanical luminosity) of the wind is lower than that of the
radiation field by a factor of order $v/c$:
\begin{equation}
\mo{v^2\over 2} = {v\over c}f\le = {2(f\le)^2\over \mo c^2}.
\label{en}
\end{equation}

\section{THE BLACKBODY COMPONENT}

Since the outflow is Compton thick for $\mo \sim \me$, 
much of the accretion luminosity generated 
deep in the potential well near $\rs$ must emerge as blackbody--like 
emission from it. If $b \sim 1$ the quasi--spherical radiating area is
\begin{equation}
A_{\rm phot} = 4\pi b\rp^2 
\end{equation}
If instead $b << 1$ the accretion luminosity emerges from the curved
sides of the outflow cones, with radiating area 
\begin{equation}
A_{\perp} = 2\pi\rp^2b^{1/2}
\end{equation}
We can combine these two cases as
\begin{equation}
A_{\rm ph, \perp} = 4\pi g\biggl[{\mo c^2\over 2\le}\biggr]^4\rs^2
\label{area}
\end{equation}
with $g(b) = 1/b, 1/2b^{1/2}$ in the two cases.
Again $g \sim 1$ unless $b < 10^{-2}$, so the areas are similar in
the two cases. However we note that the radiation patterns differ. In
particular if $b$ is small radiation is emitted over a
wider solid angle than the matter. 
Numerically we have
\begin{equation}
A_{\rm ph, \perp} = 2\times 10^{29}g\dot M_1^4M_8^{-2}~{\rm cm}^2,
\label{aeff}
\end{equation}
where $\dot M_1 = \mo/(1\msun~{\rm yr}^{-1}), M_8 = M/10^8\msun$.
The effective blackbody temperature is
\begin{equation}
T_{\rm eff} = 1\times 10^5g^{-1/4}\dot M_1^{-1}M_8^{3/4}~{\rm K}.
\label{teff}
\end{equation}
Clearly such a component is a promising candidate for the soft excess
observed in many AGN and ULXs.

\section{THE $M_{\rm BH} - \sigma$ RELATION}

It is now widely accepted that the centre of every galaxy contains a
supermassive black hole. The close observational correlation 
(Ferrarese \& Merritt, 2000; Gebhardt et al., 2000; Tremaine et al., 2002)
between the mass $M$ of this hole and the velocity dispersion $\sigma$ of 
the host bulge strongly suggests a connection between the formation of
the black hole and of the galaxy itself. 

The outflows considered above must have an important effect on this
problem. We know that most of the mass of the nuclear black holes is
assembled by luminous accretion (Soltan 1982; Yu \& Tremaine,
2002). It seems likely that the rate at which mass tries to flow in
towards the central black hole in a galaxy is set by conditions far
from the hole, for example by interactions or mergers with other
galaxies. It is quite possible therefore that super--Eddington
conditions prevail for most of the time that the central black hole
mass is being built up. Evidently the resulting Eddington thrust
(\ref{mom}) can have an important effect on the host galaxy. Unlike
luminous energy, a large fraction of a mechanical energy flux like
(\ref{en}) is likely to be absorbed within the galaxy. To reach its
present mass the black hole in PG1211+143 could have accreted at a
rate comparable to its current one for $\sim 5\times 10^7$~yr. During
that time, an outflow like the observed one could have deposited
almost $10^{60}$~erg in the host galaxy. This exceeds the binding
energy $\sim 10^{59}$~erg of a bulge with $10^{11}~\msun$ and $\sigma
\sim 300~{\rm km~s}^{-1}$.

Ideas presented by Silk and Rees (1998, henceforth SR98) and also
considered by Haehnelt et al. (1998), Blandford (1999) and Fabian
(1999) are relevant here.  These authors envisage a situation in which
the initial black holes formed with masses $\sim 10^6\msun$ before
most of the stars.  Accretion on to these black holes is assumed to
produce outflow, which interacts with the surrounding gas.  Without a
detailed treatment of the outflow from a supercritically accreting
black hole, SR98 used dimensional arguments to suggest a relation
between $M$ and $\sigma$. However this still has a free
parameter. Given the simple relation (\ref{mom}) one can now remove
this freedom. The situation turns out to resemble a momentum--driven
stellar wind bubble. Modelling this gives an $M_{\rm BH} - \sigma$
relation devoid of free parameters, and remarkably close to the
observed relation.

\subsection{Black hole wind bubbles}

SR98 modelled a protogalaxy as an isothermal sphere of
dark matter. If the gas fraction is $f_g = \Omega_{\rm
baryon}/\Omega_{\rm matter}\simeq 0.16$ (Spergel et al., 2003) its
density is
\begin{equation}
\rho = {f_g\sigma^2\over 2\pi Gr^2}
\label{rho}
\end{equation}
where $\sigma$ is assumed constant. The gas mass inside radius $R$ is
\begin{equation}
M(R) = 4\pi\int_0^R\rho r^2 {\rm d}r = {2f_g\sigma^2R\over G}
\label{m}
\end{equation}
I now assume that mass flows towards the central black
hole at some supercritical rate $\dot M_{\rm acc}$, and thus exerts a
momentum flux (\ref{mom}) on the surrounding gas, sweeping it up into
a shell.
As is well known from the theory of stellar wind bubbles
(e.g. Lamers \& Casinelli 1999) this shell is bounded by an inner shock where
the wind velocity is thermalized, and an outer shock where the surrounding
gas is heated and compressed by the wind. These two regions are
separated by a contact discontinuity. The shell velocity depends on whether
the shocked wind gas is able to cool (`momentum--driven' flow) or not
(`energy--driven' flow). In the absence of a detailed treatment of a
quasar wind, SR98 appear to have assumed the second case. In fact for the
supercritical outflows envisaged here it is easy to show that the
bubble is efficiently Compton cooled, and thus momentum--driven
instead (King, 2003)

\subsection{The $M_{\rm BH} - \sigma$ relation}

The speed $v_m$ of the momentum--driven shell now follows from the 
standard wind bubble argument. At sufficiently large radii $R$ the
swept--up shell mass $M(R)$ is much larger than the wind mass, and the
shell expands under the impinging wind ram pressure $\rho v^2$ (this
characterizes momentum--driven flows; in an energy--driven flow the
thermal pressure of the shocked wind gas is dominant, while in a
supernova blast wave the momentum injection is instantaneous rather
than continuous). The shell's equation of motion is thus
\begin{equation}
{{\rm d}\over {\rm d}t}\biggl[M(R)\dot R\biggr] = 
4\pi R^2\rho v^2 = \mo v = {\le\over c}
\label{motion}
\end{equation}
where we have used first the mass conservation equation for the quasar
wind, and then (\ref{mom}) to simplify the rhs. Integrating this
equation for $\dot R$ with the final form of the rhs gives
\begin{equation}
M(R)\dot R = {\le\over c}t
\end{equation}
where I have neglected the integration constant as $M(R)$ is
dominated by swept--up mass at large $t$. Using (\ref{m})
for $M(R)$ and integrating
once more gives
\begin{equation}
R^2 = {G\le \over 2f_g\sigma^2c}t^2,
\end{equation}
where again we may neglect the integration constant for large
$t$. We see that in the snowplow phase the shell moves with constant
velocity $v_m = R/t$, with
\begin{equation}
v_m^2 = {G\le \over 2f_g\sigma^2c}.
\label{vs}
\end{equation}

Note that this velocity is larger for higher $\le$, i.e. higher
black hole mass. This solution holds if the shell is inside the
cooling radius $R_c$; outside this radius the shell speed eventually
increases to the energy--driven value $v_e$, which also grows with
$M_{\rm BH}$.

I now consider the growth of the black hole mass by accretion.
Initially the mass is small, inflow is definitely supercritical, and
even the energy--driven shell velocity would be smaller than the
escape velocity $\sigma$. No mass is driven away, and accretion at a
rate $\dot M_{\rm Edd}$ can occur efficiently.  However as the black
hole grows, we eventually reach a situation in which $v_e > \sigma >
v_m$. Further growth is now only possible until the shell
reaches $R_c$, and then only until the point where $v_m =
\sigma$. Thus given an adequate mass supply, e.g. through
mergers, the final black hole mass is given by setting $v_m =
\sigma$ in (\ref{vs}). Thus we find the relation
\begin{equation}
M_{\rm BH} = {f_g\over 2\pi}{\kappa\over G^2}\sigma^4 \simeq
1.5\times 10^8\sigma_{200}^4~\msun. 
\label{msigma}
\end{equation}
This is remarkably close to the observed relation (Tremaine et al.,
2000). The relation also has no free parameter, unlike earlier
derivations (SR98; Haehnelt et al., 1998; Blandford 1999; Fabian
1999) If $v$ had been larger by an optical depth factor $\tau> 1$
(i.e. $R_{\rm ph}, R_{\perp}$ were outside the escape radius $R_{\rm
esc}$) a factor $1/\tau$ would have appeared on the rhs. Taking this
factor $\sim 1$ as implicitly done in deriving eqn (\ref{esc}) above
agrees with the observed outflow velocities $\sim (GM/R_{\rm
ph})^{1/2}$ in supercritically accreting quasars.

This derivation of the $M_{\rm BH} - \sigma$ relation implies that the
central black holes in galaxies gain most of their mass in phases of
super--Eddington inflow. As relatively few AGN are observed in such
phases, these must either be obscured (cf Fabian, 1999) or at high
redshift. It appears then that those quasars which are apparently now
accreting at such rates (Pounds et al., 2003a,b; Reeves et al., 2003)
are laggards in gaining mass. This idea agrees with the general
picture that these objects -- all narrow--line quasars -- are
super--Eddington because they have low black--hole masses, rather than
unusually high mass inflow rates.

\section{ULX NEBULAE}

We have seen that outflows from Eddington--limited AGN can have a
profound effect on their surroundings, in the form of the $M_{\rm BH}
- \sigma$ relation. In the stellar--mass context, ULXs are the most
supercritically accreting objects, and one would naturally expect a
similar effect here too. Indeed, many ULXs are observed to lie in the
centres of unusually large and bright emission nebulae. These have in
the past been variously interpreted as supernova/hypernova remnants
(cf Roberts et al, 2003) or photoionized nebulae (Pakull \& Mirioni,
2003). However it is clear from the work of the last Section that a
more likely explanation of the unusual energetics of these objects is
as bubbles blown in the ISM by an Eddington outflow from the
ULX. Pakull \& Mirioni (2003) indeed note that the nebulae are
probably shock excited. This interpretation means that the nebulae are
powered by the accumulated kinetic energy of the outflow that was
dissipated over the gas cooling time in the nebula (or the ULX
lifetime if this is shorter). In particular there is no requirement
that the radiation currently emitted from the ULX should power the
emission nebula. The nebulae do not provide arguments against
anisotropic emission from this source: even if the wind is
anisotropic, it is probable that it would have precessed many times
over the ULX's lifetime, and thus left a roughly spherical nebula, as
observed. One cannot therefore use ULX nebulae to argue against the
interpretation of ULXs as anisotropically emitting stellar--mass
X--ray binaries (cf King et al., 2001).

\section{CONCLUSION}

Mass outflows from Eddington--limited accreting compact objects appear
to be a very widespread phenomenon. They have major implications for
quasars and galaxy formation, and for ULXs. Some of these are barely
explored as yet, and the field promises to be fruitful.

\vspace{1pc}

I thank Andy Fabian, Jim Pringle and James Binney for illuminating
discussions. Some of the work reported here was carried out at the
Center for Astrophysics, and I thank members of its staff,
particularly Pepi Fabbiano and Martin Elvis, for stimulating
discussions and warm hospitality.  I gratefully acknowledge a Royal
Society Wolfson Research Merit Award.

\end{document}